\documentclass[aps,prb,showpacs,preprint]{revtex4}
\usepackage[dvips]{graphicx}
\usepackage{bm}
\begin{document}

\title{Cohesive and magnetic properties of grain boundaries in bcc Fe with Cr additions}

\author{E. Wachowicz, T. Ossowski, A. Kiejna}

\affiliation{Institute of Experimental Physics and Interdisciplinary Centre for Materials Modeling, University of Wroc{\l}aw, Plac M. Borna 9, PL-50-204 Wroc{\l}aw, Poland}


\begin{abstract}
Structural, cohesive, and magnetic properties of two symmetric $\Sigma3(111)$ and $\Sigma5(210)$ tilt grain boundaries (GBs) in pure bcc Fe and in dilute FeCr alloys are studied from first principles. Different concentration and position of Cr solute atoms are considered. We found that Cr atoms placed in the GB interstice enhance the cohesion by 0.5-1.2 J/m$^2$. Substitutional Cr in the layers adjacent to the boundary shows anisotropic effect on the GB cohesion: it is neutral when placed in the (111) oriented Fe grains, and enhances cohesion (by 0.5 J/m$^2$) when substituted in the boundary layer of the (210) grains. The strengthening effect of the Cr solute is dominated by the chemical component of the adhesive binding energy. Our calculations show that unlike the free iron surfaces, Cr impurities segregate to the boundaries of the Fe grains. The magnetic moments on GB atoms are substantially changed and their variation correlates with the corresponding relaxation pattern of the GB planes. The moments on Cr additions are 2-4 times enhanced in comparison with that in a Cr crystal and are antiparallel to the moments on the Fe atoms.
\end{abstract}


\maketitle

\section{Introduction}

Iron and steels have been used by mankind for four thousand years but our knowledge of their properties is still incomplete. The mechanical properties of macroscopic  polycrystalline iron are to much extent governed by cohesion at grain boundaries which, in turn, is highly dependent on the local atomic structure. 
Even the purest iron obtained in technological processes contains enough impurities \cite{Horn83} to affect the structure and chemistry of interfaces on atomic level, when segregated to the GB. Impurities may have either detrimental or beneficial effect on the GB cohesion. The former is manifested in the intergranular embrittlement (decohesion) and the latter in the strengthening of the material. 

The composition and structure of GB can be determined experimentally by the high resolution transmission electron microscopy and the x-ray diffraction methods. However, it is quite difficult to measure accurate data on the interface thermodynamic quantities. Thus \textit{ab initio} quantum mechanical methods based on the density functional theory (DFT) provide the most appropriate tool to obtain reliable quantitative information on GB structure and energetics on an electronic level. 

First principles DFT calculations of intergranular cohesion in iron in the presence of segregated impurities, using supercell models of GBs, were pioneered by Krasko and Olson.\cite{KraOls90,KraOls91} They were followed by very extensive calculations by Freeman, Olson and co-workers, \cite{WuFO94a,TangFO94,WuFO94,WuFO96,ZhoWFO97,GengFWO,GenFO01,GenFOls01,KimGF04}  who considered several different impurities or the alloying elements segregated at the $\Sigma3$ (111) symmetrical tilt GB \cite{note}
in bcc iron, by means of the full potential linearized augmented-plane-wave (FP LAPW) method. More recently the effect of impurities at that boundary was studied by using the projector augmented wave (PAW) approach.\cite{YamSK06,YamNK07} 
To our knowledge there are only few \textit{ab initio} calculations for other boundaries in Fe. The properties of the $\Sigma5$ Fe(210) GB with several nonmagnetic impurities were studied using different exchange-correlation density functionals.\cite{BraRez05,WacKie08} Results for the $\Sigma 5$(310) GB doped with Si and Sn were also reported.\cite{CakSH08}
Besides, impurity segregation and co-segregation at the $\Sigma 3$(111),  \cite{SagOE98,FenW01} and $\Sigma5$(010) boundaries \cite{ZheW01} were calculated from first principles using atomic cluster geometries.
As demonstrated by the semiempirical tight-binding calculations, \cite{YesNAPY98} ferromagnetism of iron plays a stabilizing role in intergranular cohesion. However, the magnetism at Fe GBs was not extensively explored from first principles. 
Hampel \textit{et al.} \cite{HamVC93} studied a pure and isolated $\Sigma5$(310) GB, and reported an enhanced magnetic moment at the two layers adjacent to the unrelaxed boundary. The variations in the magnetic moments at relaxed GBs in iron doped with different impurities were discussed accordingly for the $\Sigma 3$(111), \cite{WuFO94,WuFO96,ZhoWFO97,KimGF04,SagOE98} $\Sigma5$(210),\cite{WacKie08} and $\Sigma5$(310). \cite{CakSH08} In all cases the magnetic moments at the GB were  substantially enhanced and showed a damped oscillatory decrease towards the bulk value. 

Iron and chromium form a perfect solid solution which is ferromagnetic to quite low concentrations of iron. Both Fe and Cr are basic components of ferritic martensitic steels \cite{Horn83} and find many useful applications. These motivate intensive studies on the FeCr system. First principles calculations have been extensively used to study structural properties and to describe the electronic structure effects such as competition between ferro- and antiferromagnetism in the FeCr alloy.  \cite{OlsAW06,KlaDF06,OlsDW07,KlaOF07,PaxF08}
These calculations have provided a lot of important information on the mixing behavior and the heat of formation of various FeCr alloy structures with small ($\sim$10\%) Cr contents, \cite{KlaDF06} about interactions of Cr impurities with point defects in bcc Fe, \cite{OlsDW07} and the energetics of interstitials in the bulk FeCr alloy systems. \cite{KlaOF07} However, to the best of our knowledge, so far the effect of Cr additions on the cohesion at the iron GBs has not been studied from first principles. 

In this work we address the effect of low concentration of the solute Cr atoms on the GB properties in ferromagnetic $\alpha$-Fe. The properties of such a dilute FeCr alloy are affected by a complex interplay between magnetism and different structural settings of both constituents. By means of the total energy calculations we investigate the relationships between the interfacial structure and the corresponding energetic, electronic and magnetic properties at the GBs in dilute FeCr alloys. Two symmetric tilt GBs, $\Sigma5$(210) and $\Sigma 3$(111), were chosen to study the effect of concentration of the magnetic, alloying additions on cohesion/decohesion of iron boundaries, and the effect of anisotropy and the reduced coordination at the GB on the magnetic properties of the systems. 

In the next section we describe some details of our DFT calculations and define the energetic quantities which are used in the analysis and discussion of our results presented in Section \ref{sec:results}. In Section \ref{sec:summary} there is a summary.

\section{Methods of calculation}

We performed total-energy calculations based on the DFT which exploit the iterative solution of the Kohn-Sham equations in a plane-wave basis set. \cite{vasp1,vasp1a,vasp2,vasp2a} Plane waves with a kinetic-energy cutoff of 350 eV were included into calculations which yielded well converged results. The electron-ionic core interactions were described by the projector-augmented-wave (PAW) in the implementation of Kresse and Joubert. \cite{paw} The PAW method \cite{Bloechl} combines the accuracy of all-electron methods and the computational simplicity of the pseudopotential approach. 
The exchange-correlation energy was treated in the spin-polarized generalized gradient approximation (GGA) using the PW91 parametrization. \cite{GGA_PW91}

\begin{figure}
\includegraphics[width=8 cm]{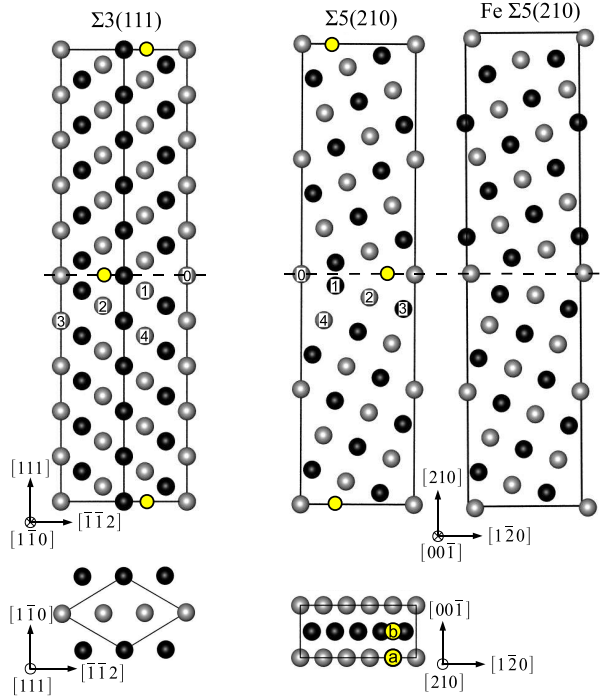}
\caption{(Color online) Side view of the supercells representing $\Sigma3$(111) and $\Sigma5$(210) boundaries between grains in bcc Fe. The lighter and darker balls mark the atoms belonging to two different planes. The right-hand side panel shows the layers stacking in the clean, relaxed Fe $\Sigma 5$(210) slab. The open circles indicate the positions of the Cr addition at the GB interstice. The substitutional positions in different grain layers are labeled by the numbers. Lower panels show top view of the unrelaxed 1$\times$1 supercells taken in the cross-section plane passing through the GB (broken line). The `\texttt{a}' and `\texttt{b}' on the $\Sigma5$ top-view panel label two different sites in the GB interstice. They can be related to the octahedral sites in a bcc unit cell with site `\texttt{a}' placed in the middle of the edge formed by two adjacent \{001\} faces, and site `\texttt{b}' at the center of the \{001\} face.}
\label{f1_cell_view} 
\end{figure} 

The 70.5$^\circ$ $\Sigma 3$ and  53.1$^\circ$ $\Sigma 5$ tilt grain boundaries are created by cutting out from the bcc crystal respectively, the (111) or (210) oriented slab of several atomic layers representing the grain and making it in contact with its image mirrored with respect to the GB symmetry-plane (Fig.~\ref{f1_cell_view}). The system is repeated periodically in space thus forming two antiparallel GBs per supercell. The (111) and (210) oriented grains were built respectively of 15 and 20 Fe atomic layers. In constructing the grains we used the theoretical equilibrium lattice parameter, $a= 2.844$ \AA, of the ferromagnetic bcc Fe, determined by us previously \cite{BloKie07} within GGA, in a good agreement with a measured value (2.867 \AA).
The slabs used in the calculations for GBs consisted of two grains, and were large enough to eliminate the spurious interaction between the two boundaries present in the supercell. 
The reciprocal space was sampled with the $8\times 8\times 1$ and $4 \times 8 \times 1$ special $k$-point meshes. \cite{MonP76} In the calculation of the fractional occupancies we applied the first order Methfessel-Paxton \cite{MethP89} method of the Fermi surface smearing with a width of 0.2 eV.  In order to find the optimum grains placement, with respect to each other, the volume and shape of the supercell representing the GB were relaxed, and all atoms were allowed to optimize their positions until the forces on each atom converged to less than 0.05 eV/\AA. 
After relaxation of the GB system, the slabs representing free surfaces were created by removing the atoms representing the second grain. Thus, for the free surface (FS), the size and shape of the supercell were adopted from the GB calculations and kept frozen, while positions of all the grain atoms were relaxed. 

\begin{figure}
\includegraphics[width=7.5 cm,bb=92 19 427 403]{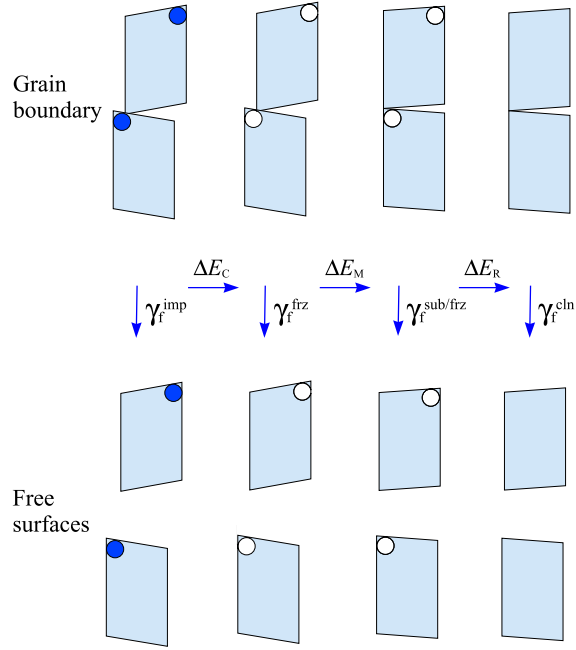}
\caption{(Color online) Schematic diagram illustrating the slab configurations used to analyze different contributions to the GB adhesive energy for a substitutional impurity atom. Upper panels show respectively (from left to right) the GB slabs: (a) the GB relaxed with an impurity atom, (b) the GB with frozen host atoms while the impurity is removed, (c) the clean GB with frozen atoms and a hole for impurity insertion, and (d)  the relaxed clean GB. The lower panels show the corresponding slabs representing the free surfaces. Dark blue balls represent impurity while the white ones mark an empty space remaining after either the impurity or host atom is removed.}\label{fig:en_schem}
\end{figure}

In order to discuss cohesive and mechanical properties of GBs it is convenient to define \cite{BraRez05} the grain boundary adhesive binding (formation) energy as
\begin{equation}
 	\gamma_{\rm f} = E_{\rm GB}  - 2E_{\rm FS}, \label{eq_1}
\end{equation}
where $E_{\rm GB}$ is the total energy of the grains at their equilibrium positions with respect to each other and $2E_{\rm FS}$ is the total energy of the two (infinitely separated) free surfaces which form the GB, taking into account all relaxation processes. 
For two identical grains in full registry the $\gamma_{\rm f}$ is (negative of) twice the surface energy. This quantity is useful in determining the effect of the solute-induced embrittlement based on a thermodynamic approach of Rice and Wang. \cite{RicW89} 
The key quantity that determines the strengthening or embrittling effect of an impurity is the strengthening energy,\cite{KimGF04}  $\Delta E_{\rm SE}$. Within the \textit{ab initio} approach it can be defined \cite{BraRez05} as the difference between the energy of binding of an impurity to the GB, $\Delta E_{\rm GB}= E_{\rm I/GB} -E_{\rm GB} -2E_{\rm I}$, or to the FS slab, $\Delta E_{\rm FS}= E_{\rm I/FS} -E_{\rm FS} -E_{\rm I}$, where $E_{\rm I/GB(FS)}$ is the total energy of the GB (or FS) system with an impurity, and $E_{\rm I}$ is the total energy of an isolated impurity. Thus, the strengthening energy can be written as \cite{WacKie08}
\begin{equation}
 \Delta E_{\rm SE}= \Delta E_{\rm GB} - 2\Delta E_{\rm FS} 
 = \gamma_{\rm f}^{\rm imp} - \gamma_{\rm f}^{\rm cln}.    \label{eq_2_SE}
\end{equation}
Here, $\gamma_{\rm f}^{\rm imp}$ is the adhesive binding energy of the GB with an impurity, and $\gamma_{\rm f}^{\rm cln}$ is the respective energy of the clean GB. A positive/negative value of $\Delta E_{\rm SE}$ means that an impurity weakens/strengthens the GB. 

The weakening/strengthening of a GB due to the presence of impurities is predominantly caused either by the chemical effect due to the electronic charge redistribution or by a structural size effect connected with a mechanical distortion of the system. There is no perfect way for an unambiguous separation of the two effects. In our analysis we follow the approach proposed by Lozovoi \textit{et al.},\cite{LozPF06} according to which the adhesive binding energy change caused by a presence of a substitutional impurity can be decomposed into the \textit{chemical}, \textit{mechanical} and \textit{host removal} energy contributions. Using the nomenclature explained in Fig. \ref{fig:en_schem}, the different energy components can be defined as follows:\\
(i) The chemical component:
\begin{equation}
  \Delta E_{\rm C} = \gamma^{\rm imp}_{\rm f} -\gamma^{\rm frz}_{\rm f} , \label{eq_3_chem}
\end{equation}
where $\gamma^{\rm frz}_{\rm f}$ and $\gamma^{\rm frz}_{\rm f}$ are respectively, the adhesive binding energies of the GB with an impurity, and of the GB with the atomic positions frozen in the relaxed GB configuration calculated with the impurity, but now with the impurity removed. \\
(ii) The mechanical contribution, $\Delta E_{\rm M}$, which accounts for the energy release during the host atoms relaxation resulting from the impurity insertion  
\begin{equation}
 \Delta E_{\rm M} = \gamma^{\rm frz}_{\rm f} -\gamma^{\rm sub/frz}_{\rm f} , \label{eq_4_mech}
\end{equation}
where $\gamma^{\rm sub/frz}_{\rm f}$ is the adhesive binding energy calculated for the clean GB frozen in the relaxed configuration, and with a removed host atom replaced by the impurity.\\
(iii) The energy change, $\Delta E_{\rm R}$, resulting from the removal of a host atom: 
\begin{equation}
 \Delta E_{\rm R} = \gamma^{\rm sub/frz}_{\rm f} - \gamma^{\rm cln}_{\rm f} . \label{eq_5_rem}
\end{equation} 
For an interstitial impurity there is no removal of host atom and thus the mechanical contribution is given by: 
\begin{equation}
  \Delta E_{\rm M} = \gamma^{\rm frz}_{\rm f} - \gamma^{\rm cln}_{\rm f}. \label{eq_6_M}
\end{equation} 

In analogy to the calculations for free surfaces \cite{Pon07,KieWac08} the segregation energy (enthalpy) of solute atom at the host GB can be calculated as the following total energy difference 
\begin{equation} \label{eq_8_segr}
	E_{\rm segr}=E_{\rm Cr,GB} - E_{\rm Cr,bulk} ,
\end{equation}
where $E_{\rm Cr,GB}$ and $E_{\rm Cr,bulk}$ are the total energies of the slab with one of the host atoms, respectively at the GB or in the bulk, substituted by the Cr. The negative $E_{\rm segr}$ means that impurity segregates at the GB.

\section{Results and discussion}\label{sec:results}
\subsection{Grain boundaries in pure Fe}\label{sec:clean}

The application of the relaxation procedure described above allows to find optimal volume and interlayer distances in the examined systems. The optimal excess volume of the relaxed grains was determined from the change in the supercell height which resulted from the relaxation of the atomic layer positions. The latter is defined as the percentage change in the vertical positions of the atoms of two subsequent atomic layers, $i$ and $j$, in a crystallite with respect to the interplanar distance in the bulk crystal, $d$, and is given by $\Delta_{ij}=[(d_j - d_i) - d]/d$.
Relaxation of the ionic and the supercell degrees of freedom may cause a parallel shift of the grains in the boundary plane which turns a symmetric tilt boundary into an asymmetric one. \cite{WacKie08}

\begin{figure}
\includegraphics*[width=7.0cm]{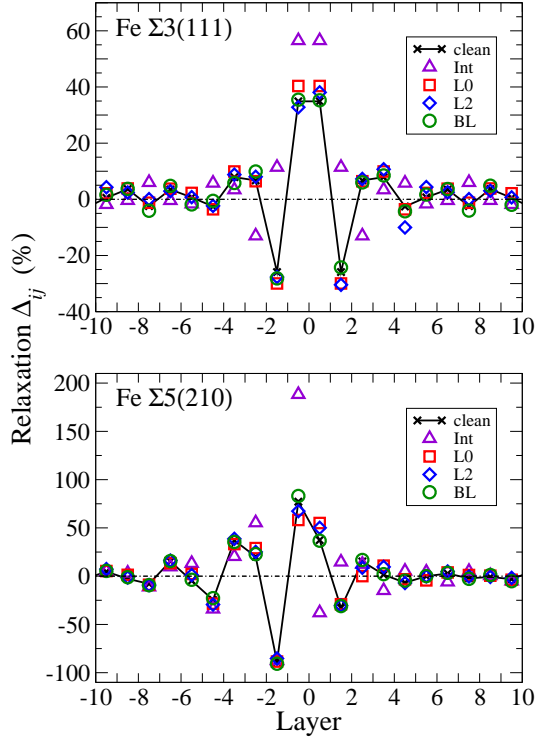}
\caption{(Color online) Relaxations of the interplanar spacing in the Fe grains near the boundary. The (111) and (210) interplanar distances in the bulk truncated Fe are 0.822 and 0.636 \AA, respectively. Also shown is the effect of high (areal) concentration of the solute Cr atoms (cf.\ section \ref{subsec:geometry}) placed in the GB interstice (Int) or in the substitutional sites of different layers (L) across the GB slab. The BL denotes the bulk Fe layer. } 
\label{f2_relax_fe}
\end{figure}  

The calculated relaxations of the interplanar distance at the $\Sigma 3$(111) and $\Sigma 5$(210) GBs in iron are displayed in Fig. \ref{f2_relax_fe}. The relaxations are very large for the first two planes [up to 35\% expansion ($+$) for $\Delta_{1,2}$ at the $\Sigma 3$ and 80\% for the $\Sigma 5$ GB, and $\sim$30\% and 90\% contraction ($-$) for $\Delta_{2,3}$, respectively] and show an oscillatory ($+-++$) decay pattern towards the bulk layers. For the pure Fe $\Sigma 3$ (111) oriented grains the maximum relaxation is doubled compared to that of the free Fe(111) surface,\cite{BloKie07} and agree well with previous FP LAPW calculations.\cite{WuFO96,ZhoWFO97,GengFWO}
These large relaxations result in a 0.23 \AA\ increase of the excess volume per unit area, in comparison to the ideal grains, which means expansion of the space available for the impurity element. 
No grain shift in the directions parallel to the $\Sigma 3$ GB is observed and consequently the relaxation pattern remains symmetric with respect to the GB plane (Fig. \ref{f2_relax_fe}).
In contrast, the (210) oriented Fe grains are substantially shifted in the GB plane in the $[1\overline{2}0]$ direction. The magnitude of the grains shift (0.6 \AA) is in line with that reported previously.\cite{BraRez05,WacKie08} 
This grains-shift enhances even more the vertical interlayer relaxations in the $\Sigma 5$ grains (up to about 80\%). It means that they are more than tripled compared to the free Fe(210) surface.\cite{BloKie07} Consequently, at the Fe $\Sigma 5$(210) boundary the optimum grains excess volume per unit area (grains separation) is further increased to 0.29 \AA\ (for the $1\times 2$ supercell) and 0.24 \AA\ (for the $1\times 1$ cell). 
For another, the $\Sigma 5$(310) GB, recent work reported \cite{CakSH08} relaxations of the interplanar distance of up to $\sim$24\%, which is about 2/3 of that for the free (310) surface.\cite{BloKie07} 
Apparently, there is a correlation between the magnitude of relaxations and coordination of the GB atoms. The coordination in the surface layer, and the surface density of atoms of the three surfaces, is decreasing in the following order: (310), (111), and (210), which means that the (210) surface is most open. This shows a clear trend: the more open the surfaces forming the boundary are the larger enhancement of GB relaxation is observed. 

\begin{table}
\caption{Calculated adhesive binding energy, $\gamma_{\rm f}$, for the pure  $\Sigma3(111)$ and $\Sigma5(210)$ boundaries in iron. Results for the GB energy, $\gamma_{\rm gb}$, which determines cohesive properties of the grains, are also presented.  The latter is defined as the difference between the total GB energy $E_{\rm GB}$ and the sum of the energies of equivalent number ($n$) of the bulk Fe atoms: $\gamma_{\rm gb} = E_{\rm GB} - n E_{\rm atom}^{\rm bulk}$.} \label{tab1:form_en_clean}  
\begin{ruledtabular}
\begin{tabular}{@{}lcccc} 
Boundary & \multicolumn{2}{c}{$\Sigma3(111)$} & \multicolumn{2}{c}{$\Sigma5(210)$} \\ 
       & (J/m$^2$) & (eV/atom)  & (J/m$^2$) & (eV/atom) \\ 
\cline{2-3}\cline{4-5} 
 $\gamma_{\rm f}$   & -3.78  & -3.27 & -3.19 & -3.49   \\
 $\gamma_{\rm gb}$ & 1.57  &  1.36  &  2.00 &  2.22   \\
\end{tabular}
\end{ruledtabular}
\end{table}

In discussing mechanical properties of GBs and the effect of additions, using Eqs.~(\ref{eq_1})-(\ref{eq_6_M}), we will compare the adhesive binding energies, $\gamma_{\rm f}$, for the clean GBs (Table \ref{tab1:form_en_clean}) with those for the GBs with Cr additions. The energies per atom (Table \ref{tab1:form_en_clean}) calculated using a small $1\times 1$ and a larger cell ($2\times2$ for the $\Sigma 3$, and $1\times2$ for the $\Sigma 5$)  agree within 0.01 eV, which gives a rough estimation of the accuracy of our calculations. For the $\Sigma 5$(210) GB the value of $\gamma_{\rm f}$ agrees well with that determined by us previously \cite{WacKie08} within GGA, and is about 2/3 of the value calculated within LDA.\cite{BraRez05} This points to the importance of a proper description of the electron exchange-correlation effects in quantification of the GB energetics.

In order to see how the presence of GBs weakens the metallic cohesion one can compare the (average) cohesive energy in the crystal with GB (enthalpy of GB formation) and that of the ideal ferromagnetic Fe crystal. Calculated as the total energy difference of the bcc Fe crystal and that of the isolated Fe atoms,  one gets that the former is lower by 0.03 and 0.05 eV/atom for the $\Sigma 3$ and $\Sigma 5$ GB, respectively, than that of the ideal Fe crystal (5.11 eV/atom).
A better measure of the cohesive strength provides the GB energy, $\gamma_{\rm gb}$, presented in Table~\ref{tab1:form_en_clean}. The anisotropy ratio of the GB energies per unit area equals to 1.27, which is still larger than the anisotropy ratio observed for the surface energies of the respective FS facets, \cite{BloKie07} and can be linked to a substantial reconstruction of the $\Sigma 5$(210) boundary. 
The GB energy of the $\Sigma 3$(111) agrees well both with previous DFT calculations \cite{YamSK06} and recent molecular dynamics study. \cite{ShiTS09} It is also of similar magnitude as that (1.63 J/m$^2$) of the $\Sigma 5$(310). \cite{CakSH08} The values of the $\gamma_{\rm gb}$ give approximately 56\% and 73\% of the energy of the free (111) and (210) surfaces,\cite{YamSK06,BloKie07} and thus they confirm the well-known correlation between the GB energy and the one-half to two-third fraction of surface energy value.\cite{YamSK06}

\begin{figure}
\includegraphics*[width=7.0cm]{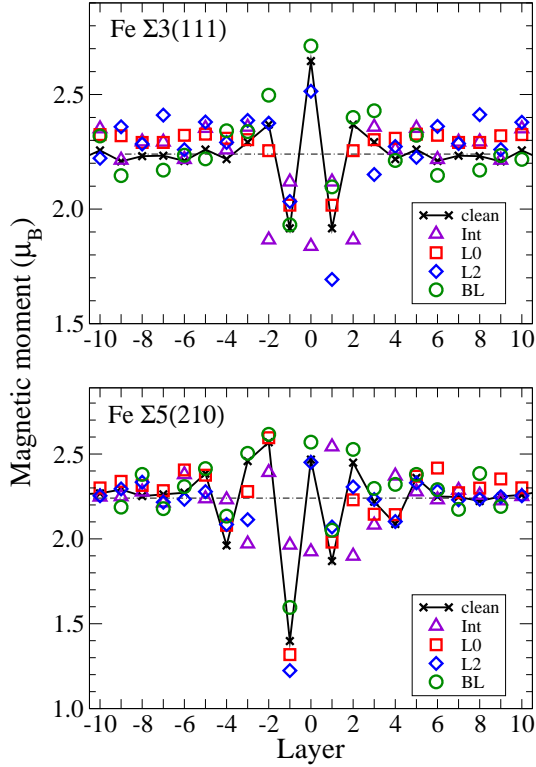}
\caption{(Color online) Magnetic moment on Fe atoms at various layers in the vicinity of the GB (cf.\ Fig.~\ref{f1_cell_view}). Horizontal chain line marks the magnetic moment of the bulk Fe. The effect of a monolayer of the solute Cr atoms (cf.\ section \ref{subsec:mm}) placed in the interstitial (Int) or substitutional sites of different layers (L) is shown as well.} \label{fig3:Fe_mag-mom}
\end{figure}  

The calculated local magnetic moments on Fe atoms ($M_{\rm Fe}$) of particular layers in the vicinity of GBs are displayed in Fig~\ref{fig3:Fe_mag-mom}. As it is seen, at the clean Fe interfaces the $M_{\rm Fe}$ can be either much increased or decreased, compared to the interior of the fully relaxed grains, depending on whether the coordination of the GB atom is improved or worsen. The moments exhibit a damped oscillatory convergence towards 2.24~$\mu_B$, when going to the deeper grain-layers. The latter value compares well with the 2.20 $\mu_B$ which characterizes the bcc Fe crystal.\cite{BloKie07} The oscillations correlate with those observed in the relaxations of the interplanar distance in the Fe grains (cf.\ Fig.~\ref{f2_relax_fe}). 
The magnitude and pattern of the $M_{\rm Fe}$ variation at the $\Sigma3$(111) GB agree very well with those reported previously. \cite{WuFO96,YesNAPY98}
The $M_{\rm Fe}$ on the boundary plane atom is increased by 15-18\%, to reach 2.65 $\mu_B$ and is followed by a similar-size decrease ($\sim$17\%), to 1.92 $\mu_B$, in the next Fe layer.
At the $\Sigma5$(210) the variations in the local $M_{\rm Fe}$ are greatly influenced by the grains shift which makes the oscillatory variation of the moment asymmetric with respect to the GB plane. 
The largest enhancement of the local $M_{\rm Fe}$ at the $\Sigma5$ GB is by 16\%, whereas the largest reduction (by 37\%) occurs next to the GB plane (1.4 $\mu_B$). This is about the same as the enhancement reported for the $\Sigma5$(310) where the $M_{\rm Fe}$ has reached 2.55 $\mu_B$. \cite{CakSH08} This may suggest that the moments are equal when the number of coincidence sites at two GBs, which are indicated by $\Sigma$, are equal. Also it seems that the moments are smaller when the inverse density of coincidence sites is higher (i.e., a higher $M_{\rm Fe}$ for a lower $\Sigma$).
The local $M_{\rm Fe}$ on the GB-plane atoms decreases in the same order as does the coordination in the surface layers of different grain facets which differs from the free Fe surfaces where they are ranked as $M_{\rm Fe}(310) > M_{\rm Fe}(210) > M_{\rm Fe}(111)$. However, it should be noted that the $M_{\rm Fe}$ of the FSs shows a smaller anisotropy and all moments are within 2.81--2.88 $\mu_B$.\cite{BloKie07}

\subsection{Cr impurities at Fe grain boundaries}

By placing one impurity atom in the GB cells of different size, we examined two different Cr concentrations at each GB: a monolayer and a quarter of monolayer of Cr in the $1\times 1$ and $2\times 2$ supercells of the $\Sigma 3$ GB, and a monolayer and a half a monolayer of Cr in the $1\times 1$ and $1\times 2$ cells at the $\Sigma 5$. Note that throughout this work, when discussing dependencies on higher and lower Cr concentration, we mean  accordingly, a high (a monolayer) and a low \textit{areal} concentration of the solute atoms. In all considered cases the average volume concentration of Cr was smaller than 7\%.  

\subsubsection{Geometry and cohesion.}\label{subsec:geometry}

Impurities usually modify the positions of host atoms and influence the relative positions of the grains. The atomic radius of Cr is 1.27 \AA\ and is very close to that of Fe (1.25~\AA). Thus, Cr impurity substituted to the Fe matrix should not introduce any substantial strain to the host structure. For a high (areal) Cr concentration considered by us, the substitution means the replacement of one of the whole Fe layers by Cr.

The changes in the interlayer relaxations caused by a monolayer of Cr substituted for different Fe layers are shown in Fig.~\ref{f2_relax_fe}. 
For clarity only the effect of impurities placed in the layers adjacent to the boundary and deep inside the grain is shown. As it can be seen, the changes are limited to the GB region. In general, relaxations in Fe are only little affected by the presence of substitutional Cr.  At the $\Sigma 3$ GB the solute atoms increase, while at the $\Sigma 5$ boundary they suppress, or do not alter, the relaxations. For the $\Sigma 3$ GB, the relaxations become slightly asymmetric if Cr is not situated exactly at the interface plane. At the $\Sigma 5$ GB, the biggest changes ($\sim 20$~\%) with respect to the relaxed clean GB, appear for Cr replacing exactly the boundary Fe layer (L0). The size and relaxation pattern caused by the substitutional Cr is altered only slightly when calculated (not shown) for the lower Cr concentration ($2\times2$ and $1\times2$ cells for the $\Sigma 3$ and $\Sigma 5$, respectively).

In contrast to substitutional Cr additions, a monolayer of Cr placed in the GB interstice (Fig.~\ref{f2_relax_fe}) increases meaningfully interlayer relaxations in the grains (up to 60\% and 180\%, at the $\Sigma 3$(111) and $\Sigma 5$(210) GBs, respectively). Besides, its influence on the interplanar relaxations extends over a wider region than in the case of substitutional Cr. 
At the relaxed $\Sigma 5$(210) GB in bcc Fe there are two types of the interstitial holes (Fig.~\ref{f1_cell_view}), located in the interstice between the neighboring (001) planes of the two grains, which are convenient for impurity atom placement. 
Cr atom placed in the hole of type `\texttt{a}' binds to three Fe atoms in the same (001) layer (lighter balls in Fig.~\ref{f1_cell_view}) with the bond lengths of 1.3--1.8 \AA. The \texttt{b}-type hole which is formed in the neighboring (001) plane, represented by the darker balls in Fig.~\ref{f1_cell_view}, is coordinated by nine Fe atoms -- three atoms from the same layer and three more atoms from each of the neighboring planes -- with the bond lengths ranging between 1.9--2.5 \AA. 
For a monolayer of Cr the two places are equally favorable (within 24 meV). However, for a lower areal concentration of Cr in the \texttt{a}-sites, the total energy of the supercell is by 0.89 eV lower than that with Cr in \texttt{b}-sites. Therefore, in further calculations only site \texttt{a} was considered. 

\begin{figure}
\includegraphics*[width=8.5 cm]{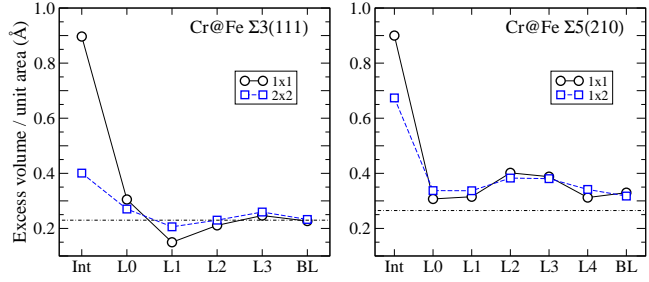}
\caption{(Color online) Change in the excess volume per unit area of the Fe-grains caused by Cr impurity placed in the GB interstice (\texttt{Int}) or substituted in different Fe layers (L) across the boundaries (cf.\ Fig.\ref{f1_cell_view}). The BL labels bulk (central) layer of the grain with Cr atom (L7 for the $\Sigma 3$(111), and L9 for the $\Sigma 5$(210) oriented slabs). The dashed lines mark the calculated excess volume for the clean GBs.}  
\label{fig4:change_separ}
\end{figure} 

The changes in the relaxation generally lead to an increased excess volume (Fig.~\ref{fig4:change_separ}). A distinct exception is when Cr is placed in the L1 layer at the $\Sigma 3$ GB. For Cr substituted in one of the layers of the $\Sigma 5$ GB the excess volume is always enhanced, compared to the clean GB and is largest for the Cr placed in the second or third Fe layer. Cr addition placed either in the $\Sigma 3$ or $\Sigma 5$ GB interstice causes a much greater increase in the excess volume of the Fe grains (Fig.~\ref{fig4:change_separ}). 
The excess volume per unit area increment can reach $\simeq 0.9$~\AA, for a monolayer of Cr. The magnitude of the excess volume is similar for the two GBs considered, but it results from different changes in the relaxation pattern. 
Recently, it was found \cite{BraRez05,WacKie08} that a monolayer of the nonalloying impurity inserted substitutionally at the Fe $\Sigma 5(210)$ GB may cause a very large shift of the grains. Present calculations confirm those findings for the alloying element and show that the solute atoms placed in the $\Sigma5$ GB interstice, in concentrations corresponding to the full monolayer and half a monolayer of Cr ($1\times 1$ and $1\times 2$ cells), cause the grains shift of 2.32 and 2.35 \AA, respectively. 
These shifts are a consequence of creating a new layer by Cr atoms which should be shifted by $\sim$2.5 \AA, according to both the layers stacking order and the initial shift of the pure grains. In case of the substitutional Cr the size of the shift is not altered with respect to that for the relaxed clean GB ($\simeq$0.6 \AA). No grains shift is found for Cr atoms appearing at the $\Sigma3$(111) GB. 

\begin{figure}
\includegraphics*[width=8.5 cm]{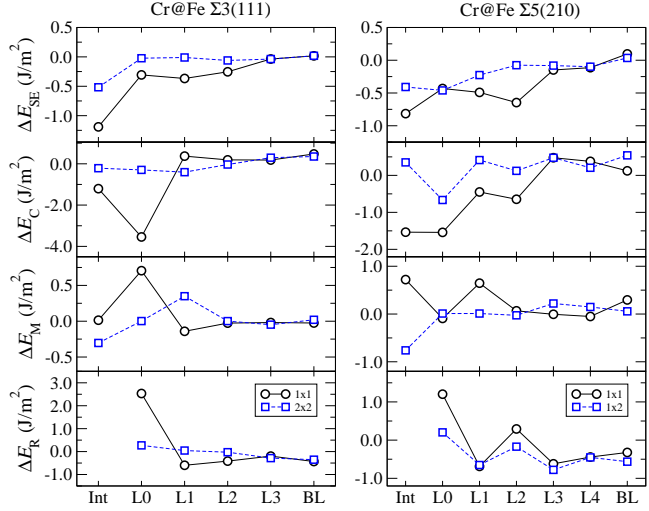}
\caption{(Color online) Strengthening energy (top panels) and its chemical, mechanical, and host removal energy components for two GBs in iron doped with the Cr impurity placed accordingly in the interstitial or substitutional sites across the GB slab. All energies are in J/m$^2$.} 
\label{fig5:decomp_111}
\end{figure} 

Calculated energies of the GB strengthening caused by the solute atoms are displayed in Fig.~\ref{fig5:decomp_111}. Smaller concentration of the substitutional Cr does not actually influence the adhesive binding energy of the $\Sigma3$ GB. This agrees quite well with a small weakening (of 0.02 eV/atom) reported by Geng \textit{et al.}\cite{GenFO01} who considered the effect of substitutional alloying on the GB cohesion in Fe, within the electronic-level phenomenological theory based on the first-principles calculations. A monolayer of Cr, in turn, enhances cohesion if  substituted for one of the first three layers of the $\Sigma3$ GB. At the $\Sigma5$ GB a small strengthening ($\sim$0.1 J/m$^2$) of the GB cohesion is observed for the Cr monolayer placed as deep as in L4. 
Considering that the interlayer distance in the (210) oriented grain is smaller than that in the (111), the region where the Cr substitution has any influence on the GB properties is $\sim$2.5 \AA\ thick in both cases. But similarly to Cr, at the Fe $\Sigma3$(111) the cohesion enhancement is more pronounced for the higher Cr concentration. The interstitial Cr is a distinctly stronger cohesion enhancer at Fe boundaries than the substitutional Cr (Fig.~\ref{fig5:decomp_111}), even at lower concentration. For a monolayer of Cr at the GB interstice the cohesion enhancement exceeds 1 J/m$^2$. 

A decomposition of the strengthening energy into different contributions to the cohesion (Fig.~\ref{fig5:decomp_111}) shows that for a lower concentration of the interstitial Cr at the  ${\Sigma}$3 GB, the cohesion enhancement is mainly due to the mechanical component, whereas for a high Cr concentration the chemical effects dominate. 
For substitutional Cr, the energy contributions are small when the Cr atoms are placed in the third (L2) or deeper layers of the grain, regardless their concentration. A monolayer of Cr situated in the boundary layer (L0) induces strong chemical interactions. 
A relatively large chemical-energy component is due to the electron charge redistribution. Calculated changes in the electron charge distribution (not shown) demonstrate that there is a charge density increase, mainly in the vicinity of the Cr atom, when Cr atom is situated in the GB region. This contributes to a stronger bonding between Cr and the neighboring Fe atoms and leads to an enhancement of the chemical contribution. For Cr placed deep inside the grains no meaningful electron density change in the boundary region is observed and the chemical component remains practically unchanged. 
The chemical energy, however, is to much extent compensated by the mechanical and the host removal energy components. In contrast, when the Cr replaced Fe atoms of L1 or L2, the cohesion enhancement results mainly from the lowering in mechanical and host removal energy components. 

At the ${\Sigma}$5 GB the variations in the energies are noticeable for the Cr atom placed as deep as in the fourth boundary layer (Fig.~\ref{fig5:decomp_111}). For a lower Cr concentration the mechanical component is close to zero, and  the observed cohesion enhancement results from an interplay between the chemical and the host removal energy components. This is in contrast to the interstitial impurity, where the cohesion enhancement is due to the lowering in the mechanical component. For a monolayer of Cr the chemical interaction is a main reason for the observed cohesion enhancement for all  impurity placements. 

\begin{figure}
\includegraphics*[width=8.5 cm]{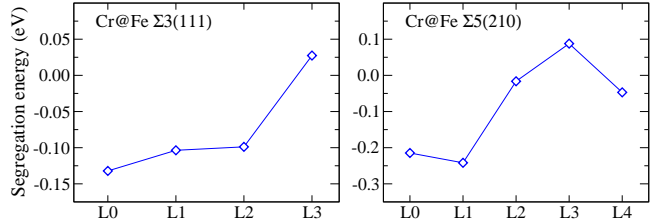}
\caption{(Color online) Energy of segregation of the Cr solute in different layers at the $\Sigma 3$(111) and $\Sigma 5$(210) boundaries in iron for small Cr concentration.} 
\label{fig6_segreg_Cr_in_Fe}
\end{figure} 

\subsubsection{Cr segregation at Fe GBs.}

Surface energy of the free Fe-surfaces \cite{BloKie07} is distinctly smaller than that of the Cr facets.\cite{OssKie08} Thus, according to simple thermodynamical  argument, Cr should not segregate to the free Fe surface. Also recent first-principles calculations of the segregation of Cr at several FSs of the dilute FeCr alloys,\cite{Pon07,KieWac08} have shown that segregation energy is positive, which means that Cr should not enrich the surface of the dilute FeCr system.\cite{KieWac08}
It was also shown, however, that Cr may segregate at the Fe surface for higher bulk concentration of Cr. On the other hand, situation might be different at the GB where the coordination of atoms differs both from that in the bulk of the grain and that at the free surface. 

In order to check the segregation behavior of Cr at the Fe GBs we compared the total energy of the system with a single Cr atom placed in one of the Fe-layers adjacent to the boundary and that with Cr atom in the middle (bulk) layer of the grain [cf.~Eq.~(\ref{eq_8_segr})]. 
Calculated segregation energies (Fig.~\ref{fig6_segreg_Cr_in_Fe}) show that Cr atoms exhibit a clear tendency to enrich the GBs in Fe. 
The same concentrations of Cr, both per volume of the grain and per area of the boundary plane, applied in this work as well as in our previous work on free surfaces of the dilute FeCr alloy,\cite{KieWac08} allow for a direct comparison of the segregation behavior of the two systems. 
In contrast to the free (111) and (210) iron surfaces where the enrichment by Cr was found to be unfavored (by $\sim$0.2--0.3 eV),\cite{KieWac08} regardless the position of Cr solute atom in the surface or subsurface Fe layers, at the $\Sigma3$(111) and  $\Sigma5$(210) GBs segregation is favorable in the first three layers of the Fe grains. The segregation of Cr is most pronounced in the first two layers closest to the boundary plane, where the energy of the enriched $\Sigma3$ and $\Sigma5$ GB is lowered approximately by $-$0.1 and $-$0.2 eV, respectively. For the $\Sigma 5$ boundary there exists a barrier for Cr segregation at the third layer, where segregation becomes unfavorable. 
Interestingly, we also found that having a monolayer of Cr at the GB (not shown) is more favored (by $-0.26$ eV at the $\Sigma 3$, and $-0.31$ eV at the $\Sigma 3$ GB) than to have it in the middle of the grain. It is worth noting that at the dilute concentration, Fe atoms placed in the chromium grains \cite{OsoWK09} preferably enrich the interface Cr layers.

\begin{figure}
 \includegraphics*[width=8.5cm]{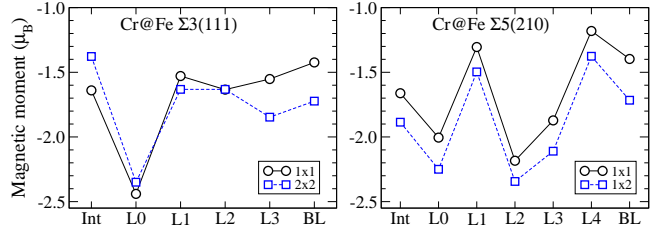}
 \caption{(Color online) Magnetic moment on the Cr solute atom placed at the GB interstice or substitutionally in different layers at two different  boundaries in iron.} \label{fig7:Cr_in_Fe_mag-mom}
 \end{figure} 

\subsubsection{Magnetic properties.}\label{subsec:mm}

Figure~\ref{fig3:Fe_mag-mom} shows variations in the magnetic moment on atoms of the Fe grains in the direction perpendicular to the GB, induced by a monolayer of Cr. When the solute atom is placed near the ${\Sigma}$3 GB, the moments on the Fe atoms near the boundary are reduced by $\sim$5\% and show a larger amplitude of variations in the deeper layers. An asymmetric placement of the impurity with respect to the GB plane is reflected in the antisymmetric changes of the $M_{\rm Fe}$. For the lower Cr concentration (not shown) the changes in the moments are smaller. At the ${\Sigma}$5 GB the amplitude of the $M_{\rm Fe}$  variations is generally increased compared to the pure boundary but in some layers the directions of variation are reversed, depending on the situation of the Cr atom. 

The variations in the magnetic moment on Cr impurity ($M_{\rm Cr}$) placed in different sites in the two differently oriented Fe grains are shown in Fig.~\ref{fig7:Cr_in_Fe_mag-mom}. A minus sign of the $M_{\rm Cr}$ means that it is coupled antiparallel to the moments on the neighboring Fe atoms. Independently of the orientation of the grains, deep in the Fe interior the $M_{\rm Cr}$ attains $-$1.4 and $-$1.7 $\mu_B$, for a monolayer of Cr and a lower Cr concentrations, respectively. These values are much higher than 0.59 $\mu_B$ which is characteristic for the bulk bcc Cr crystal.\cite{OssKie08}  
They agree very well with the results reported by Klaver \textit{et al.}\cite{KlaDF06} for the interior of the dilute FeCr alloys. For a single Cr atom in the bulk Fe they obtained the magnetic moment of $\sim -1.75 \mu_B$, and for a Cr atom with one additional Cr in the neighborhood $\sim-1.5 \mu_B$. This may be an indication of the fact that for smaller Cr concentration in the grains the Cr-Cr repulsive interaction resulting from magnetic frustration is very small or even negligible. 
As one can see the behavior of the magnetic moment at the two GBs is different. While at the $\Sigma3$ GB the local $M_{\rm Cr}$ varies rather moderately, between $-$1.4 and $-$1.7 $\mu_B$, except the case of Cr at the boundary plane (L0) where it achieves $-2.4$ $\mu_B$, at the $\Sigma5$(210) GB the variations span the range from $-1.2$ to $-2.4$ $\mu_B$, and are largest on the Cr placed substitutionally. At lower Cr concentration (1 Cr per $1\times 2$ cell) the moment on Cr in the very boundary plane (L0), or in the second layer (L2) of the grain, approaches the values of $-$2.2 and $-$2.4 $\mu_B$, respectively, i.e., it is equal or even exceeds that on Fe atoms in the interior of Fe crystal. 
At the ${\Sigma}5$ GB the changes of the $M_{\rm Cr}$ are very alike for the two examined Cr concentration and differ by $\sim$0.3 $\mu_B$ which can be attributed to the Cr-Cr repulsion resulting from the magnetic frustration. 

\begin{figure}
\includegraphics*[width=8.5 cm]{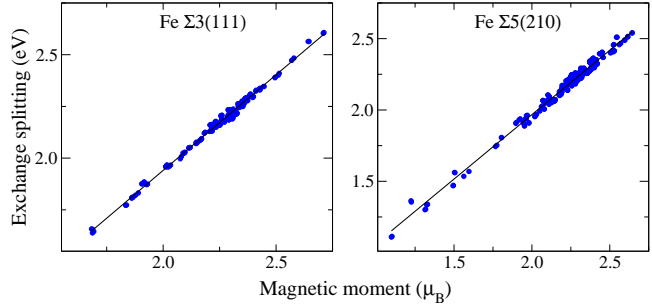}
\caption{Correlation between the magnetic moment and the exchange splitting for the Fe atoms at the $\Sigma3$(111) and $\Sigma5$(210) GBs. The data points represent all examined cases of the clean and doped GBs, calculated in the $1\times 1$ supercell. The magnetic moment on the bulk Fe atom is 2.24 $\mu_B$.}
\label{fig8:exch_split_Fe}
\end{figure} 

The magnetic properties of GBs atoms are very sensitive to their local environment. In the Stoner theory \cite{Sto38} of itinerant magnetism the origin of a ferromagnetic order is explained by a rigid shift of the spin-up and spin-down bands under the influence of the exchange interaction. The ratio of this exchange splitting to the local magnetic moment (the Stoner parameter) plays an essential role in determining magnetic behavior of the system. Its empirical value was found to be quite universal and close to 1 eV/$\mu_B$ for a wide class of magnetic systems.\cite{Him91}
Recently, it was shown that exactly the same linear correlation holds also for the atoms of Fe grains around the clean $\Sigma5$(310) boundary.\cite{CakSH08} On the other hand, earlier theoretical studies have reported a slightly lower proportionality ratio of 0.92 to 0.95 eV/$\mu_B$ for the Fe-atom sites in the intermetallic systems.\cite{Gun76,BeuHEF94,TurHaf92}
Calculated as the difference in the centers of gravity of the local density of states for spin-up and spin-down electrons \cite{CakSH08,TurHaf92} the Stoner parameter for Fe atoms at the distorted GB can be plotted versus the local magnetic moments on the respective atoms  (Fig.~\ref{fig8:exch_split_Fe}). The respective dependencies for the $\Sigma3$ and $\Sigma5$ GB-atoms, can be approximated by the linear least-squares fits $y = 0.90 \mu_B$ + 0.16,  and $y= 0.92 \mu_B + 0.09$. The scatter of the data around the straight line (Fig.~\ref{fig8:exch_split_Fe}) is larger for the $\Sigma5$ which is more distorted than the $\Sigma3$(111).
Similar relations, though with slightly smaller slopes (0.88 eV/$\mu_B$ and 0.90 eV/$\mu_B$ for the $\Sigma3$ and $\Sigma5$ GBs, respectively) hold also when moments on Cr solute atoms are included. Thus, at the GBs in iron with Cr additions the magnetic order is primarily ruled by the magnetism in Fe and can be described by Stoner's model of ferromagnetism. 

\section{Summary}\label{sec:summary}

In this work we have investigated from first principles the structural, cohesive, and magnetic properties of two high-angle tilt grain boundaries in iron, the $\Sigma3$(111) and $\Sigma5$ (210), with a small amount of Cr additions. 
At clean the GBs the interplanar distances in the grains are greatly enhanced. Full relaxation of the system caused a substantial parallel shift of the $\Sigma5$(210) grains, while the $\Sigma3$ GB remained unreconstructed. 
The formation of the clean $\Sigma3$ GB, where one-in-three GB atoms coincide,  costs more energy than to create the $\Sigma5$ with one-in-five-atoms coincidence. The magnetic moments on Fe atoms at the GB exhibit an oscillatory variation with the atomic layer depth. 
The chromium additions placed substitutionally do not change much the relaxation pattern, whereas interstitial Cr increases greatly the interlayer distances, and consequently enhances the grains excess volume per unit area up to four times. 
A monolayer of the substitutional Cr enhances cohesion at the $\Sigma3$ GB, while at smaller concentration it is neutral. Added at the $\Sigma5$ GB chromium strengthens the GB. 
A decomposition of the strengthening energy shows that responsible for the cohesion enhancement are the changes in the bonding at the GB. Both for Cr substituted into the Fe matrix and the interstitial Cr, the chemical contribution, resulting from the electron charge redistribution, dominates alone and causes the GB  strengthening. Placed at the GB interstice, Cr has a beneficial effect on the cohesion, the strengthening being stronger for a monolayer concentration. 
We have shown that unlike the free Fe surfaces, the  enrichment of iron GBs by dilute Cr is energetically favorable.  
The magnetic moment on the host Fe atoms is generally reduced when the Cr-solute atoms are present while the moment on the Cr-solute is much increased compared to that in a Cr crystal. It is demonstrated that the magnetic order at the GBs in iron (both clean and with Cr additions) can be explained by the Stoner model relating the local magnetic moments to the amount of exchange-splitting in the bands.

\acknowledgments
We appreciate useful discussions with M.\ W.\ Finnis and K.\ J.\ Kurzyd{\l}owski. This work was supported by the Polish Ministry of Science and Higher Education under Grant No.\ COST/201/2006. We acknowledge allocation of computer time  by the Interdisciplinary Centre for Mathematical and Computational Modeling (ICM), University of Warsaw (Grant G28-25).



\begin{thebibliography}{49}
\bibitem{note} Throughout this work we apply the coincidence site lattice notation where $\Sigma$ denotes the reciprocal density of coincidence sites of atomic lattice.\cite{SutBal95}
\end{thebibliography}
\end{document}